\documentclass[prl,twocolumn, showpacs,floatfix,amsfonts]{revtex4}
\usepackage{graphicx,graphics,color,epsfig}
\usepackage{bm}
\usepackage{amsmath}
\usepackage{amssymb}

\newcommand{\bk}{{\bf k}}

\newcommand{\br}{{\bf r}}

\renewcommand{\Re}{{\rm Re}}

\begin{document}
\preprint{}
\title{Noise spectroscopy and interlayer phase-coherence in bilayer quantum Hall systems}
\author{Yogesh N. Joglekar$^1$, Alexander V. Balatsky$^1$ and Allan H. MacDonald$^2$}
\affiliation{$^1$Theoretical Division, Los Alamos National Laboratory, Los Alamos, New Mexico 87544.\\
$^2$Department of Physics, University of Texas at Austin, Austin, Texas 78712.}
\date{\today}

\begin{abstract}
Bilayer quantum Hall systems develop strong interlayer phase-coherence when the distance between layers 
is comparable to the typical distance between electrons within a layer. The phase-coherent state has until 
now been investigated primarily via transport measurements. We argue here that interlayer current and 
charge-imbalance noise studies in these systems will be able to address some of the key experimental 
questions. We show that the characteristic frequency of current-noise is that of the zero wavevector 
collective mode, which is sensitive to the degree of order in the system. Local electric potential noise 
measured in a plane above the bilayer system on the other hand is sensitive to finite-wavevector collective 
modes and hence to the soft-magnetoroton picture of the order-disorder phase transition.
\end{abstract}
\pacs{73.21.-b}
\maketitle

{\it Introduction:}
Bilayer quantum Hall systems near total filling factor $\nu=1$ undergo a quantum phase transition from a 
compressible state to an incompressible state with interlayer phase-coherence as the ratio of layer separation 
to magnetic length $d/l$ is reduced~\cite{murphy}. This phase-coherent state survives in the limit of 
vanishing interlayer tunneling $\Delta_t$, in which case the charge-gap in the phase-coherent state arises 
purely because of electron-electron interactions between the two layers. The incompressible state can be 
regarded either as an easy-plane ferromagnet~\cite{fertig,abp,kmky,qhreviews} or as excitonic 
superfluid~\cite{elhole}. The estimated tunneling amplitude in the experimental samples, 
$\Delta_t\sim 50\,\mu$K, is much smaller than any other energy-scale~\cite{ibs}, suggesting that interlayer 
phase-coherence is established spontaneously when the ratio $d/l$ is smaller than some critical value 
$d_{cr}/l$. Transport measurements in these systems change spectacularly when the 
compressible-to-incompressible phase boundary is crossed~\cite{ibs,mk} and can largely be explained in terms 
of excitonic condensate and quasiparticle contributions~\cite{bilayertheory,prl}. However, key questions 
remain about both the nature of the phase transition and the mechanism which drives it. Is the transition 
between a phase-coherent state and an exotic compressible state?~\cite{demler} Is the transition driven by the 
percolation of regions which are incompressible?~\cite{ady} Is the transition weakly first-order in the 
absence of disorder?~\cite{john} Is the transition associated with softening of the collective-mode at a 
finite wavevector?~\cite{fertig,abp,pel} How does disorder affect these scenarios? To address these questions, 
it is necessary to understand the evolution of the two important energy scales, the collective-mode gap at 
zero wavevector and the roton-minimum gap near $kl\sim 1$, near the phase-boundary. In this Letter we show 
that noise-spectroscopy of interlayer current probes the collective mode energy at zero wavevector, whereas 
noise-spectroscopy of charge imbalance probes the roton-like minimum in the collective mode energy. Combined, 
these two measurements can establish the nature of the transition and shed light on effect of disorder on 
the transition.

Noise analysis has been used extensively to probe intrinsic properties of many different 
systems~\cite{noiserefs}. For example, it can be used to measure the atomic magnetic resonance in a sample 
without an external noise source~\cite{mitsui} and has been suggested as a technique for single-spin detection 
using a scanning tunneling microscope~\cite{jhu}. In this paper we propose that the same technique can be used 
to answer key questions in bilayer quantum Hall systems. Although the average current between layers is zero 
when the system is not driven by an interlayer electrochemical potential difference, spectral analysis of 
fluctuations in the current can reveal the condensate and quasiparticle energy scales of the phase-coherent 
state. Current fluctuations of the system are controlled by the current-current susceptibility. In the absence 
of an electron-hole condensate, this susceptibility is controlled by the incoherent contribution of 
quasiparticles in the system, and the correlation time of fluctuations will be dominated by the quasiparticle 
energy gap. In contrast, the appearance of the condensate implies a collective mode which transfers charge 
from one layer to the other. This collective mode will contribute to the fluctuations; hence in the ordered 
state we will have a qualitatively different spectrum of fluctuations. A similar analysis applies to the 
fluctuations in the charge imbalance which lead to fluctuations in the electrostatic potential. These effects 
are the basis of the noise spectroscopy we propose. We find that the current-current correlator oscillates 
with characteristic frequency equal to the collective-mode energy at zero wavevector, 
$\omega_c=E_{sw}({\bf 0})$ (we use units such that $\hbar=1$), if the interlayer transport is predominantly 
collective. We also find that the electrostatic-potential correlator at long times has a characteristic 
frequency equal to the collective-mode energy minimum, $\omega^{*}=E_{sw}^{*}$, which occurs at a finite 
wavevector $kl\sim 1$. Thus noise-spectroscopy of a bilayer system can probe the two important energy scales, 
$E_{sw}({\bf 0})$ and $E_{sw}^{*}$, in the phase-coherent state and their evolution as $d\rightarrow d_{cr}$.

In the following paragraphs, we describe the microscopic model we use for a biased bilayer system, obtain 
analytical expressions for the current-current and electrostatic-potential correlators in terms of bilayer 
response functions, and discuss the implications of these results. 

{\it Microscopic Model:}
Let us consider a biased bilayer system with filling factor $\nu_T$ in the top layer and $\nu_B$ in the 
bottom layer such that the total filling factor $\nu=\nu_T+\nu_B=1$. We use a pseudospin language to describe 
the bilayer system.
\begin{equation}
\label{eq: one}
\hat{S}_\alpha(\br)=\frac{1}{2}\sum_{\sigma',\sigma}c^\dagger_{\sigma'}(\br)\tau^{(\alpha)}_{\sigma'\sigma}
c_{\sigma}(\br)
\end{equation}
is the pseudospin density in direction $\alpha$ at position $\br$, $\sigma,\sigma'$ are the pseudospin labels, 
and $\tau^{(\alpha)}_{\sigma'\sigma}$ are Pauli spin matrices with pseudospin up/down representing electrons 
in top/bottom layers. The microscopic Hamiltonian for the bilayer system is given by 
$\hat{H}=\hat{H}_0+\hat{V}$. The one-body term is 
\begin{equation}
\label{eq: two}
\hat{H}_0= -\int_{\br}\left[ \Delta_t \hat{S}_x(\br) + \Delta_v\hat{S}_z(\br)\right]
\end{equation}
where $\Delta_t$ is the interlayer tunneling amplitude and $\Delta_v$ is the bias voltage which leads to 
unequal filling factors in the two layers. The interaction term $\hat{V}$ is a sum of the intra-layer Coulomb 
interaction, $V_A(\bk)=2\pi e^2/\epsilon k$ and the interlayer Coulomb interaction $V_E(\bk)=e^{-kd}V_A(\bk)$. 
Note that the antisymmetric combination $V_x=(V_A-V_E)/2$ is not rotationally invariant in the pseudospin 
space. In the pseudospin-language, the phase-coherent incompressible state is an $XY$ easy-plane ferromagnet 
and has a uniform non-zero value of $\langle\hat{S}_z(\br)\rangle$ when the filling factors in the two layers 
are unequal~\cite{biased,tutuc}. When $\Delta_t\neq 0$ the U(1) symmetry in the $x$-$y$ plane in 
pseudospin-space is broken explicitly, and the uniform ordered moment $\vec{M}$ lies in the $x$-$z$ plane, 
$\vec{M}=M_0(\sin\theta_v,0,\cos\theta_v)$~\cite{biased}. $M_0$ is the dimensionless order parameter which 
vanishes near the incompressible-state phase-boundary and approaches 1 as the layer separation 
$d\rightarrow 0$. The angle $\theta_v$ is determined by self-consistent Hartree and exchange contributions to 
the tunneling amplitude and bias voltage, and is given by $\cos\theta_v=\Delta_v/\Delta_{vc}$ in the limit 
$\Delta_t\rightarrow 0$. Here $\Delta_{vc}(d)=2\left[V_x(0)-\Gamma_x(0)\right]$ is inverse of the 
exchange-enhanced interlayer capacitance, $\Gamma_x=(\Gamma_A-\Gamma_E)/2$, and $\Gamma_{(A/E)}(\bk)$ are the 
intra-layer/interlayer Coulomb exchange interactions.

{\it Current Noise:} The only term in the microscopic Hamiltonian which does not conserve particle number 
within each layer separately is proportional to the interlayer tunneling amplitude $\Delta_t$. Therefore the 
total interlayer current $\hat{I}$, obtained by using the continuity equation~\cite{cav1}, is given by 
\begin{equation}
\label{eq: three}
\hat{I}= e\Delta_t\int_\br\hat{S}_y(\br).
\end{equation}
In the absence of any electrochemical potential difference between the two layers the average interlayer 
current is zero, $\langle\hat{I}\rangle=0$. However, as we will see, the current-current correlations are 
nonzero. Let us consider the symmetrized current-current correlator, 
$C_{I}(t)=\langle \hat{I}(t)\hat{I}(0) + \hat{I}(0)\hat{I}(t)\rangle$. Using Eq.(\ref{eq: three}), it is 
straightforward to express the correlator in terms of pseudospin response function
\begin{equation}
\label{eq: four}
C_{I}(t)=\left(e\Delta_t\right)^2 2\Re \chi_{yy}(\bk=0,t)
\end{equation}
where $\chi_{yy}(\bk,t)\equiv\langle T\hat{S}_y(\bk,t)\hat{S}_y(-\bk,0)\rangle$ is the {\it time-ordered} 
pseudospin susceptibility. For quantum Hall bilayers, it is possible to calculate the susceptibility 
analytically from the microscopic Hamiltonian using different approximations which capture either the 
quasiparticle or the collective-mode physics. Since the interlayer transport in the phase-coherent state is 
{\it presumably} collective~\cite{ibs,mk,bilayertheory,prl}, we evaluate the susceptibility using generalized 
random phase approximation (GRPA) which captures the physics of collective modes~\cite{fertig,kmky},
\begin{equation}
\label{eq: five}
\chi_{yy}(\bk,i\Omega_n)=\frac{e^{-k^2l^2/2}}{8\pi l^2}
\frac{2iM_0 a_{\theta_v}(\bk)}{(i\Omega_n)^2-E^2_{sw}(\bk)},
\end{equation}
where $i\Omega_n$ is a bosonic Matsubara frequency and the exponential prefactor arises from projection onto 
the lowest Landau level. The collective-mode dispersion is given by 
$E_{sw}(\bk)=\sqrt{a_{\theta_v}(\bk)\cdot b_{\theta_v}(\bk)}$ where~\cite{biased}
\begin{equation}
\label{eq: six}
a_{\theta_v}(\bk)=\Delta_{qp}-M_0\Gamma_E(\bk)+2M_0\left[V_x(\bk)-\Gamma_x(\bk)\right]\sin^2\theta_v
\end{equation}
is the cost of charge-imbalance fluctuation and $b_{\theta_v}(\bk)=\Delta_{qp}-M_0\Gamma_E(\bk)$ is the cost 
of phase-fluctuations in the $x$-$y$ plane. Here, the quasiparticle energy splitting is given by 
$\Delta_{qp}=M_0\Gamma_E(0)$ when $\Delta_t\rightarrow 0$ or $\Delta_v\rightarrow 0$; for a nonzero tunneling 
and nonzero bias voltage the quasiparticle energy splitting is obtained by solving the self-consistent 
Hartree-Fock mean-field equations numerically~\cite{biased}.

We emphasize that the pseudospin susceptibility $\chi_{yy}$ is calculated in the presence of finite 
interlayer tunneling. Therefore the cost of uniform phase-fluctuations is non-zero, 
$b_{\theta_v}({\bf 0})\neq 0$, and the collective mode energy at the origin is nonzero, 
$E_{sw}({\bf 0})\sim\sqrt{\Delta_t\left(\Delta_t+M_0\Delta_{vc}\sin^2\theta_v\right)}\neq 0$. When 
$\Delta_t\rightarrow 0$, the static uniform susceptibility (\ref{eq: five}) diverges as expected in a state 
which possesses a spontaneous broken symmetry. It is necessary to treat the interlayer tunneling 
non-perturbatively because the interlayer current operator $\hat{I}$ vanishes at 
$\Delta_t=0$~\cite{prl,fertig2}. It follows from Eq.(\ref{eq: five}) that the current-current correlator is 
given by 
\begin{equation}
\label{eq: seven}
C(t)=\left(e\Delta_t\right)^2\frac{2 M_0 a_{\theta_v}({\bf 0})}{8\pi l^2E_{sw}({\bf 0})}
\cos\left(E_{sw}({\bf 0})|t|\right)
\end{equation}
Eq.(\ref{eq: seven}) is the first principle result of the paper. {\it It shows that the current-current 
correlator oscillates with characteristic frequency $\omega_c=E_{sw}({\bf 0})$} and its strength is 
proportional to $M_0$. Therefore the noise spectroscopy of interlayer current will be able to probe the 
collective mode energy at zero wavevector. Note that near the incompressible-to-compressible phase boundary, 
the only parameter in $E_{sw}({\bf 0})$ which varies rapidly with $d/l$ is the order-parameter $M_0$; the 
tunneling amplitude $\Delta_t$ and the exchange-enhanced inverse capacitance $\Delta_{vc}$ are relatively 
constant. Therefore, the dependence of the condensate frequency $\omega_c$ on $d/l$ will provide a 
{\it direct} measurement of the phase-coherent order parameter $M_0$. For typical values of system parameters, 
$\Delta_t\sim 50\,\mu$K and $\Delta_{vc}\sim 50$K, and balanced bilayers ($\theta_v=\pi/2$), we get 
$\omega_c\sim\sqrt{M_0}$ GHz (for example, with $M_0\sim 10^{-1}-10^{-2}$, we have $\omega_c\sim 100-300$ MHz).

\begin{figure}[htbp]
\begin{center}
\includegraphics[width=3.2in]{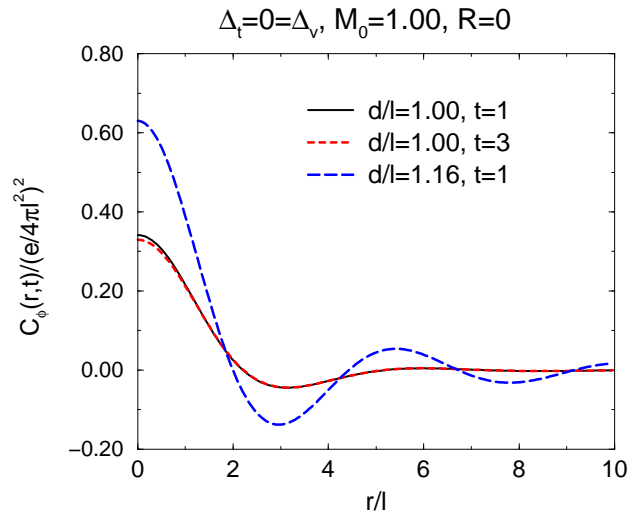}
\caption{Typical potential correlator $C_\phi(r,t)$. Time is measured in units of inverse Coulomb energy. 
$C_\phi(r,t)$ decays with $r$ due to the Bessel function $J_0(kr)$ in the integrand (\ref{eq: nine}), and is 
not very sensitive to the softening of the roton-like minimum near $kl\sim 1$ which occurs at 
$d_{cr}/l=1.18$~\cite{fertig,abp,kmky}.}
\label{fig: c_vs_r}
\vspace{-8mm}
\end{center}
\end{figure}

{\it Electrostatic Potential Noise:} Now let us consider the effect of random charge-imbalance fluctuations 
on the electrostatic potential. We assume that the potential is measured (using, for example, a 
single-electron transistor) in a plane located at distance $R$ above the top layer~\cite{yacobi}. The change 
in the electrostatic potential $\delta\phi$ produced by a change in the interlayer charge-imbalance 
is~\cite{caveat2}
\begin{equation}
\label{eq: eight}
\delta\phi(\bk)=2eV_x(-\bk) e^{-kR}\delta S_z(\bk)
\end{equation}
where $\delta S_z(\br)$ is the {\it fluctuation} in the charge-imbalance at position $\br$. In steady state, 
the potential fluctuations average to zero $\left<\delta\phi\right>=0$. However, the potential correlator is 
non-zero. Let us define 
$C_{\phi}(r,t)=\left<\delta\phi(\br,t)\delta\phi({\bf 0},0)+\delta\phi({\bf 0},0)\delta\phi(\br,t)\right>$. 
It follows from Eq.(\ref{eq: eight}) that this correlator is related to the time-ordered pseudospin 
susceptibility $\chi_{zz}$,
\begin{equation}
\label{eq: nine}
C_{\phi}(r,t)=4e^2\int\frac{kdk}{2\pi}J_0(kr)e^{-2kR}V_x^2(\bk)2\Re\chi_{zz}(\bk,t).
\end{equation}
Eq.(\ref{eq: nine}) is the second principle result of this paper. Following the analysis in the current-noise 
case, the GRPA susceptibility which captures the physics of collective excitations is given by 
\begin{equation}
\label{eq: ten}
\chi_{zz}(\bk,t)=\frac{e^{-k^2l^2/2}}{8\pi l^2}\frac{M_0b_{\theta_v}(\bk)}{E_{sw}(\bk)}
\exp\left(-iE_{sw}(\bk)|t|\right).
\end{equation}
We note that the static uniform $\chi_{zz}$ susceptibility vanishes in the limit $\Delta_t\rightarrow 0$, in 
contrast to the static uniform $\chi_{yy}$ susceptibility (\ref{eq: five}). It is also worth emphasizing that 
the potential correlator $C_{\phi}(r,t)$ can be evaluated perturbatively in the tunneling amplitude 
$\Delta_t$, whereas the current-current correlator $C_{I}(t)$ requires a non-perturbative calculation. First 
let us concentrate on the short-time behavior of the correlator function $C_{\phi}(r,t)$. It follows from 
Eq.(\ref{eq: ten}) that the correlator saturates at $t=0$ and falls off quadratically with $t$ at short times 
$|t|\ll\Delta_{qp}^{-1}$. It is primarily sensitive to the phase-fluctuations cost $b_{\theta_v}(\bk)$, which 
does not change qualitatively as a function of increasing $d/l$. Figure~\ref{fig: c_vs_r} shows the typical 
$r$-dependence of the potential correlator $C_\phi(r,t)$ at short-times for different values of $d/l$. 

To simplify the discussion of the long-time correlator, we will specialize to a bilayer system with 
$\Delta_t\rightarrow 0$ and consider only local correlations, $r=0$, although the results obtained here are 
valid in general. We remind the Reader that the approximation $\Delta_t=0$ is justified when evaluating the 
potential correlator $C_\phi(r,t)$. Restricting to a local correlator corresponds to the usual experimental 
situation with one single-electron transistor probe~\cite{yacobi} which measures (the noise in) the 
electrostatic potential. At long times, the integral in Eq.(\ref{eq: nine}) is dominated by the region in 
momentum-space where the pseudospin-wave energy is minimum. It is known that in bilayer systems the 
collective-mode dispersion typically has two minima~\cite{fertig,abp,kmky}. The first minimum occurs at the 
origin, $\bk=0$, and the collective mode dispersion in the vicinity is linear. The second minimum occurs near 
$kl\sim 1$. The collective-mode energy at this minimum, $E_{sw}^*(d)$, reduces as the ratio $d/l$ is 
increased, due to the lowered energy cost for charge-imbalance fluctuations, 
$a_{\theta_v}(kl\sim 1)\rightarrow 0$. It follows from Eq.(\ref{eq: ten}) that the region near the second 
minimum gives a diverging contribution, $\chi_{zz} \sim 1/\sqrt{a_{\theta_v}}\rightarrow\infty$, and that 
contribution varies 
strongly with layer separation $d/l$. Thus, the long-time potential correlator $C_\phi$ has a component with 
frequency $\omega^{*}=E_{sw}^{*}(d)$ which varies rapidly with $d/l$. The strength of this feature in the 
potential correlator scales as $\sqrt{M_0/(d-d_{cr})}$. The incompressible-to-compressible phase transition 
which occurs at a critical layer separation $d_{cr}$ can be understood in terms of the collapse of this 
collective-mode energy minimum, $E_{sw}^*(d_{cr})=0$~\cite{fertig,abp}. {\it A measurement of the 
characteristic frequency $\omega^{*}$ and its dependence on $d/l$ can directly probe the collective-mode 
energy gap collapse} and shed light on whether it is instrumental to the phase transition~\cite{pel}. The 
variation of $E_{sw}^{*}(d)$ with $d/l$ can also provide an independent {\it direct} measure of the order 
parameter $M_0$, since close to the phase-boundary the minimum collective-mode energy can be approximated as 
$E_{sw}^*(d)\propto\sqrt{M_0(d-d_{cr})}$.

\begin{figure}[htbp]
\begin{center}
\includegraphics[width=3.5in]{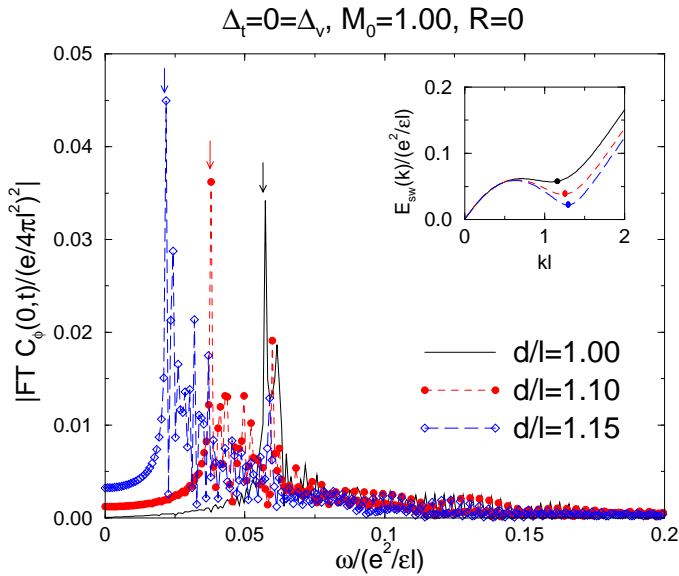}
\caption{Fourier transform (FT) of the potential correlator $C_\phi(0,t)$. The inset shows the 
collective-mode dispersions for corresponding values of $d/l$. The peak in the FT, marked by arrows, occurs 
at $\omega=E_{sw}^*(d)$ and varies with $d$ as $\sqrt{(d-d_{cr})}$; it probes the collective-mode minimum near 
$kl\sim 1$, marked by solid circles, shown in the inset.}
\label{fig: c_vs_w}
\vspace{-5mm}
\end{center}
\end{figure}

Figure~\ref{fig: c_vs_w} shows the Fourier transform of the long-time correlator $C_\phi(0,t)$ for different 
values $d/l$, whereas the inset shows the corresponding collective-mode dispersions. The peak near 
$\omega/(e^2/\epsilon l)\sim .06$ is not sensitive to $d/l$ and represents the contribution from region near 
the origin where the collective-mode dispersions for all values of $d/l$ are essentially the same. The second 
peak occurs at $\omega=\omega^{*}=E_{sw}^*(d)$ and varies rapidly with $d/l$. The strength of this frequency 
component increases when $d\rightarrow d_{cr}$ as expected from Eq.(\ref{eq: ten})~\cite{generic}. 

{\it Summary:}
We have shown that spectral analysis of fluctuations in bilayer quantum Hall systems provides a novel probe of 
the phase-coherent state. We predict that the current-noise spectrum has a characteristic frequency 
$\omega_c=E_{sw}({\bf 0})$ if the interlayer transport is collective. We also predict that the noise in the 
electrostatic potential has a dominant frequency component equal to the collective-mode energy minimum at 
finite wavevector, $\omega^{*}=E_{sw}^{*}$. These two measurements can probe directly the order-parameter of 
the phase-coherent state as well as the collapse of the collective-mode energy gap $E_{sw}^{*}$ near the 
incompressible-to-compressible phase-boundary. Our theoretical analysis of these quantities does not account 
for the influence of disorder and inhomogeneity, and does not allow for the possibility that exotic correlated 
states appear on either side of the phase transition. Although the analysis of the noise spectra in these 
scenarios is more complex than in the BCS-like excitonic condensate case considered here, it is clear that 
there will be substantial differences, especially as the phase boundary is approached. Experimental tests of 
the predictions made above, which we regard as a starting point for the interpretation of experiment, will 
shed light on the nature of the compressible-to-incompressible phase-transition, and deepen our understanding 
of the phase-coherent state in quantum Hall bilayers.


It is a pleasure to thank Z. Nussinov for useful discussions. This work was supported by DOE LDRD (LANL), by 
the NSF under grant DMR0115947 (UT), and by the Welch Foundation (UT).



\begin{thebibliography}{99}
\bibitem{murphy} S.Q. Murphy {\it et al.}, Phys. Rev. Lett. {\bf 72}, 728 (1994).
\bibitem{fertig} H.A. Fertig, Phys. Rev. B {\bfseries 40}, 1087 (1989).
\bibitem{abp} A.H. MacDonald {\it et al.}, Phys. Rev. Lett. {\bf 65}, 775 (1990).
\bibitem{kmky} K. Moon {\it et al.}, Phys. Rev. B {\bf 51}, 5138 (1995); 
K. Yang {\it et al.}, Phys. Rev. B {\bfseries 54}, 11644 (1996).
\bibitem{qhreviews} S.M. Girvin and A.H. MacDonald in {\it Perspectives in Quantum Hall Effects}, 
edited by S. Das Sarma and Aron Pinczuk; 
S. Das Sarma and E. Demler, Solid State Commun. {\bf 117}, 141 (2001).
\bibitem{elhole} Yu.E. Lozovik and V.I. Yudson, Solid State Commun. {\bf 19}, 39 (1976); 
Y. Kuramoto and C. Horie, Solid State Commun. {\bf 25}, 713 (1978); 
D. Paquet {\it et al.}, Phys. Rev. B {\bf 32}, 5208 (1985); 
A.H. MacDonald and E.H. Rezayi, Phys. Rev. B {\bf 42}, 3224 (1990); 
D. Yoshioka and A.H. MacDonald, J. Phys. Soc. Jpn. {\bf 59}, 4211 (1990).
\bibitem{ibs}I.B. Spielman {\it et al.}, Phys. Rev. Lett. {\bf 84}, 5808 (2000); 
Phys. Rev. Lett. {\bf 87}, 036803 (2001).
\bibitem{mk}M. Kellogg {\it et al.}, Phys. Rev. Lett. {\bf 88}, 126804 (2002); 
Phys. Rev. Lett. {\bf 90}, 246801 (2003).
\bibitem{bilayertheory} L. Balents and L. Radzihovsky, Phys. Rev. Lett. {\bf 86}, 1825  (2001); 
A. Stern {\it et al.}, {\it ibid.}, 1829  (2001); 
M.M. Fogler and F. Wilczek, {\it ibid.}, 1833 (2001).
\bibitem{prl} Y.N. Joglekar and A.H. MacDonald, Phys. Rev. Lett. {\bf 87}, 196802 (2001).
\bibitem{demler} E. Demler {\it et al.}, Phys. Rev. Lett. {\bf 86}, 1853 (2001).
\bibitem{ady}A. Stern and B. I. Halperin, Phys. Rev. Lett. {\bf 88}, 106801 (2002).
\bibitem{john} J. Schliemann {\it et al.}, Phys. Rev. Lett. {\bf 86}, 1849 (2001).
\bibitem{pel} S. Luin {\it et al.}, Phys. Rev. Lett. {\bf 90}, 236802 (2003).
\bibitem{noiserefs}  E.B. Aleksandrov and V.S. Zapasskii, Zh. Eksp. Teor. Fiz. {\bf 81}, 132 (1981); 
D.D. Awschalom {\it et al.}, Phys. Rev. Lett. {\bf 68}, 3092 (1992); 
Y. Manassen {\it et al.}, Phys. Rev. B {\bf 61} 16223 (2000).
\bibitem{mitsui} T. Mitsui, Phys. Rev. Lett. {\bf 84}, 5292 (2000).
\bibitem{jhu}  Y. Manassen {\it et al.}, Phys. Rev. Lett. {\bf 62}, 2531 (1989); 
Y. Manassen, J. Magn. Reson. {\bf 126}, 133 (1997); 
C. Durkan and M.E. Welland, App. Phys. Lett. {\bf 80}, 459 (2002); 
J.-X. Zhu and A.V. Balatsky, Phys. Rev. Lett. {\bf 89}, 286802 (2002); 
A.V. Balatsky {\it et al.}, Phys. Rev. B {\bf 66}, 195416 (2003); 
Z. Nussinov {\it et al.}, Phys. Rev. B {\bf 68}, 085402 (2003).
\bibitem{biased} Y.N. Joglekar and A.H. MacDonald, Phys. Rev. B {\bf 65}, 235319 (2002).
\bibitem{tutuc} E. Tutuc {\it et al.}, Phys. Rev. Lett. {\bf 91}, 076802 (2003).
\bibitem{cav1} In typical samples, current measured in the leads is the sum of local current densities. 
Therefore, we consider the correlator for the total current instead of current density.
\bibitem{fertig2} H.A. Fertig and J.P. Straley, Phys. Rev. Lett. {\bf 91}, 046806 (2003).
\bibitem{yacobi} A. Yacobi {\it et al.}, Solid State Commun. {\bf 111}, 1 (1999); 
N.B. Zhitenev {\it et al.}, Nature (London) {\bf 404}, 473 (2000).
\bibitem{caveat2} Strictly speaking, this expression is valid for an infinite sample with dielectric constant 
$\epsilon$; however, results obtained here are unchanged when we take into account the dielectric constant 
discontinuity at the sample surface. 
\bibitem{generic} These results remain valid even if the phase transition is driven by reducing the bias 
voltage~\cite{biased,tutuc} instead of increasing $d/l$.
\end{thebibliography}
\end{document}